# Towards Multidimensional Verification: Where Functional Meets Non-Functional

Maksim Jenihhin, Xinhui Lai, Tara Ghasempouri, Jaan Raik
*Computer Systems, Tallinn University of Technology, Estonia, maksim@ati.ttu.ee*

*Abstract*— Trends in advanced electronic systems' design have a notable impact on design verification technologies. The recent paradigms of Internet-of-Things (IoT) and Cyber-Physical Systems (CPS) assume devices immersed in physical environments, significantly constrained in resources and expected to provide levels of security, privacy, reliability, performance and low power features. In recent years, numerous extra-functional aspects of electronic systems were brought to the front and imply verification of hardware design models in multidimensional space along with the functional concerns of the target system. However, different from the software domain such a holistic approach remains underdeveloped. The contributions of this paper are a taxonomy for multidimensional hardware verification aspects, a state-of-the-art survey of related research works and trends towards the multidimensional verification concept. The concept is motivated by an example for the functional and power verification dimensions.

## I. INTRODUCTION

Today, several prominent trends in electronic systems design can be observed. The Internet-of-Things (IoT) and Cyber-Physical Systems (CPS) devices are immersed in physical environments, significantly constrained in resources and expected to provide levels of security and privacy [1], ultra-low power feature or high performance. Very complex electronic systems, including those built from the non-certified for reliability Commercial-Off-The-Shelf (COTS) components, are used for safety- and business-critical applications. These trends along with gigascale integration at nanoscale technology nodes and multi-/many-processor based systems-on-chip architectures have ultimately brought to the front various *extra-functional aspects* of the electronic systems' design at the chip design level. The latter include security, reliability, timing, power consumption etc. There exist numerous threats causing an electronic system to violate its specification. In the hardware part, these are design errors (bugs), manufacturing defects and variations, reliability issues such as soft errors and aging faults or malicious faults, such as security attacks. Eventually, there can also be bugs in the software part.

Hardware design model verification detects *design errors* affecting *functional* and *extra-functional* (also interchangeably referred as *non-functional*) aspects of the target electronic system. Strictly, the sole *task of extra-functional verification* of a design model is limited to detecting deviations that cause violation of extra-functional requirements. In practice, it often intersects with the task of functional verification [2], [14], thus establishing *a multidimensional space for verification*. A "grey area" in distinction between functional and extra-functional requirements may appear when an extra-functional requirement is a part of design's main functionality. E.g., security requirements for some HW design can be split into extra-functional and functional sets if the design's purpose and specified functionality is a system's security aspect, e.g. it is a secure cryptoprocessor.

In this paper, we present an overview for the recent trends in extra-functional and functional verification of HW designs and discuss the challenges towards the holistic multidimensional verification. The rest of this paper is organized as follows: Section II provides a taxonomy of multidimensional verification aspects, Sections III proposes a state-of-the-art survey with the key trends in verification for the main extra-functional aspects, Section IV discusses the multidimensional verification paradigm and presents a motivational example for the functional and power verification dimensions, Section V draws the conclusions.

## II. TAXONOMY OF MULTIDIMENSIONAL VERIFICATION ASPECTS

In practice, relevance of each functional and extra-functional aspect strongly depends on design type, target system application and specific user requirements. Following the design paradigm shift, a number of extra-functional aspects have recently received significant academic research attention e.g., security. At the same time, there already exist established industrial practices for measuring and maintaining separate design qualities, e.g. the RAS (Reliability-Availability-Serviceability) aspect introduced by IBM [6]. While in the software engineering discipline, the taxonomy of extra-functional requirements has a comprehensive coverage by the literature [7]-[12], it cannot be directly re-used for the HW verification discipline because of significant difference in the design models.

Fig 1 introduces a taxonomy of multidimensional verification aspects derived from the performed literature review. The conventional *functional concerns* are *safety and liveness properties*, *combinational and temporal dependencies* along with *data types*, however this list can be extended for particular designs. The extra-functional aspects can be strictly categorized into three groups: *system qualities, system resource constraints* and *timing aspects* (in bold). Despite the *security* and *reliability* aspects belong to the first group and the *power* aspect belongs to the second group, these three aspects have a special attention in the literature and in practical applications. Several extra-functional aspects such as *(manufacturing defects) testability*, *fault-tolerance* and other in-field fault group aspects do not have a direct correspondence in the software engineering discipline because of the distinct nature of faults. Other aspects such as *real-time constraints* are very similar between the two domains.

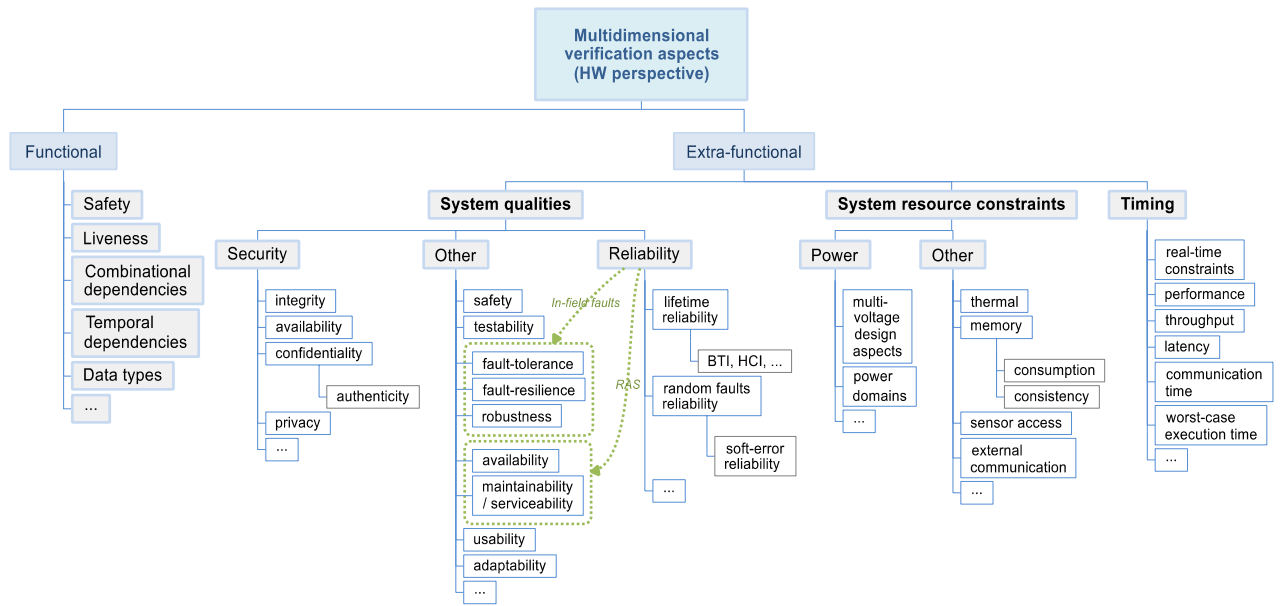

Fig. 1. Taxonomy of multidimensional verification aspects

### III. Trends in Extra-Functional Verification

Table I presents a survey of recent publications targeting extra-functional and multidimensional verification. Here, along with the specific extra-functional aspects details about the design model and verification approach are outlined, i.e., the design under verification type, verification engine, the level of abstraction, design representation language, compute model and the tool operated in the research. For instance, the row for paper [40] shows that the authors performed model checking to reduce the state space of a Timed Petri Net of a real-time scheduler. Looking at this row, real-time constraints is the type of timing property, a scheduler of an embedded system is the design under verification, the abstraction level is the system level and SMT model checker Promela [64], Timed Petri Net and SPIN [63] are the verification engine, the design representation language, the compute model and the tool, respectively. We pointed out such key points for all the recent up to 10-year old studies in this area.

#### A. Security aspects

Security is difficult to quantify as today there are no commonly agreed metrics for this purpose [1]. The key targeted security services [16] are commonly represented as non-functional aspects for verification are *confidentiality*, *integrity* and *availability*. They are tightly linked to the type of attack and the attacker model assumed for each case, i.e. black-, grey- or white-box.

Today, for complex HW designs (e.g. IEEE1687 Reconfigurable Scan Networks or NoCs) the specific on-chip security features in the design model to be verified also tend to be very sophisticated. These include on-chip mechanisms for attack prevention (firewalls, user management, communications isolation), attack protection (traffic scrambling, encryption) and attack resilience (checkers for side-channel attacks, covert channel detection, attack recovery mechanisms).

Many of the existing works in security verification (e.g. [21], [23], [25], [28], [29]) are focusing on the integrity attribute, mostly addressing HW trojan detection . There also exist some works that additionally target ([19], [20], [22], [24], [30]) or are exclusively considering ([26], [27]) the confidentiality aspect.

Several solutions in security verification are restricted to target specific architectures or types of modules such as Reconfigurable Scan Networks (RSNs) [22], [26] or macro-asynchronous micro-synchronous pipelines [29].

There is virtually no work that considers security in combination with other extra-functional aspects. Some solutions in the security verification of NoCs indirectly address reliability due to the fact that they implement hardware monitors that allow avoiding both, attacks and in-field faults [20], [21]. An approach that is designed for modeling a multitude of extra-functional aspects is the model-based engineering example of Architecture Analysis and Design Language (AADL) [19]. While, in principle, AADL allows representing several extra-functional aspects (called quality attributes in AADL), [19] only concentrates on analysis of confidentiality as a part of verifying security in a system with multiple levels of security. The authors in [70] have target a general multi-view HW modeling and verification approach taking into consideration the security view.

#### B. Reliability aspects

The key drivers for the reliability aspect in today's designs are the recent industrial standards in different application domains such as IEC61508, ISO26262, IEC61511, IEC62279, IEC62061, RTCA/DO-254, IEC60601, etc. These ultimately imply extra-functional features such as safety mechanisms and redundancy to ensure levels of fault coverage, e.g. ASIL (Automotive Safety Integrity Level). Here, the key threats are transient faults in the field such as radiation-induced single event effects or *soft errors* [15] and intermittent to permanent faults by process or time-dependent variations, i.e. *aging*, e.g. induced by Bias Temperature Instability (BTI) [13]. New applications, demand the systems to be fail-safe or fail-operational, by functionally redundant design parts enabling fault-tolerance, -resilience and -robustness. A promising initiative in reliability specification and modelling is the Reliability Information Interchange Format (RIIF) [30].

TABLE I. SURVEY OF THE STATE-OF-THE-ART SOLUTIONS FOR EXTRA-FUNCTIONAL AND MULTIDIMENSIONAL VERIFICATION

| Paper | Year[1] | Extra-functional aspect[2] | | | | | | Design under verification | Verification engine | Abstract. level[5] | Design representation language | Compute model | Tool |
|---|---|---|---|---|---|---|---|---|---|---|---|---|---|
| | | Security | Reliability[3] | Timing[4] | Power | Other system quality | Other constrained resource | | | | | | |
| [19] | 2009 | confidentiality, integrity | - | - | - | - | - | HW/SW system | formal, correct-by-construction | SL | AADL | - | OSATE |
| [20] | 2016 | integrity, confidentiality | ○ | - | - | - | - | NoC | simulation, HW monitors | RTL | VHDL/Verilog | - | - |
| [21] | 2014 | integrity | ○ | - | - | - | - | NoC | formal | GL | VHDL/Verilog | - | SurfNoC |
| [22] | 2017 | integrity, confidentiality | - | - | - | - | - | RSN | model check | RTL | ICL | Craig interpolation | CIP solver |
| [23] | 2015 | integrity | - | - | - | - | - | SoC | simulation | RTL | VHDL/Verilog | - | - |
| [24] | 2016 | integrity, confidentiality | - | - | - | - | - | ALU | equivalence check | GL | - | QBF-SAT | - |
| [25] | 2017 | integrity | - | - | - | - | - | SoC | semiformal | GL | - | - | JasperGold SPV |
| [26] | 2016 | confidentiality | - | - | - | - | - | RSN | model check | RTL | ICL | Craig interpolation | CIP Solver |
| [27] | 2017 | confidentiality | - | - | - | - | - | industrial control systems | formal | SL | ASLan++ | - | CL-AtSe |
| [28] | 2017 | integrity | - | - | - | - | - | IP cores | semiformal | GL | VHDL | - | mini-SAT |
| [29] | 2015 | integrity | - | - | - | - | - | ISA, pipeline | model check | RTL | - | CTL, LTL | nuXmv SMV |
| [30],[71] | 2013 | integrity, confidentiality | - | - | - | - | - | IPs and SoCs | formal | RTL, GL | Verilog | - | JasperGold SPV |
| [33] | 2017 | - | ● | - | - | - | - | CPS | model check | SL | AADL | Timed Automata | UPPAAL |
| [34] | 2015 | - | SER | - | - | - | - | IP cores | formal | GL/RTL | LDDL | LDDL | Coq |
| [35] | 2010 | - | SER | - | - | availability, serviceability | - | processor | fault inject. | GL | Verilog | - | IBM in-house |
| [36] | 2016 | - | ● | - | - | availability, serviceability | - | SoC | fault inject. | RTL | - | - | - |
| [38] | 2018 | - | - | latency | ○ | - | - | NoC | fault inject. | RTL | VHDL | - | QoSinNoC |
| [39] | 2011 | - | - | RT | - | - | - | memory | model check | RTL | REAL;AADL | - | Ocarina |
| [40] | 2010 | - | - | RT | - | - | - | Scheduler of a RT emb. system | model check | - | Promela | Time Petri-net | SPIN |
| [41] | 2010 | - | - | latency | - | - | - | RT emb. system | model check | SL | AADL | - | YICES |
| [42] | 2017 | - | - | performance | ○ | - | - | NoC, HW/SW architectures | simulation | SL | GAL (Graph Assembly Language) | resource / connectivity graphs | ArchOn |
| [43],[44] | 2016 | - | ○ (LTR) | - | ○ | - | thermal | Smart Systems | simulation | SL | IP-XACT; SystemC-AMS | - | - |
| [46] | 2012 | | | | ● | - | - | IPs | simulation | SL | SystemC | - | - |
| [47] | 2016 | - | - | - | ● | - | - | DSP cores | simulation | SL,GL,RTL | SystemC | - | Powersim |
| [50] | 2017 | - | - | performance | ○ | - | - | automotive CPS | model check | SL | C, EAST-ADL | Timed Automata | UPPAALsdv |
| [51] | 2016 | - | - | - | ● | - | - | IPs | semiformal ABV | RTL | VHDL/Verilog; SystemC | Hidden Markov Model | - |
| [52] | 2012 | - | - | execution time | ○ | - | - | distributed emb. system | simulation | SL | SystemC | - | - |
| [53] | 2016 | - | - | performance | ○ | - | thermal | HW/SW platform | simulation, formal (analytical) | RTL,TLM,SL | UML; C++; SystemC-AMS; VHDL | HIF | HIFSuite |
| [54] | 2014 | - | - | - | - | connectivity | - | SoC | symbolic model checking | RTL/TLM | Verilog | - | Incisive Formal Verifier |
| [55] | 2008 | - | - | ○ (latencies) | - | connectivity | - | SoC | property checking | RTL | Verilog | - | JasperGold CV |
| [56] | 2016 | - | - | - | - | memory consistency | - | processor | simulation | ISA | ruby | - | McVerSi |
| [60] | 2011 | - | - | - | ● | - | thermal | SoC | simulation | SL, GL/RTL | SystemC | - | PowerMixer, PowerDepot, PowerBrick, |
| [61] | 2015 | - | - | - | ● | - | - | | simulation | SL,TLM | SytemC | - | Power Kernel Tool |
| [62] | 2011 | - | - | - | ● | - | - | SoC | simulation | SL | SystemC | - | Powersim |
| [67] | 2018 | - | ● | - | - | - | - | CPS | formal and simulation, HW monitors | RTL | VHDL/Verilog | multiple | multiple |
| [68] | 2014 | - | SER | - | - | - | - | IPs | SAT solver | RTL | VHDL | - | - |
| [69] | 2010 | - | SER | - | - | - | - | IPs, processor | simulation | RTL | VHDL/Verilog | - | - |
| [70] | 2018 | ● | ● | - | - | - | - | MPSoC | model check | SL, RTL | - | Timed Automata | UPPAAL |

[1] only conference, journal and industrial white papers published in the last 10 years were selected for this survey
[2] ● – this aspect is the main focus in the paper; ○ – this aspect is partially addressed
[3] LTR – lifetime reliability; SER – soft-error reliability;
[4] RT – real-time constraints;
[5] GL – gate level; SL – system level; ISA – instruction set architecture level; TLM – transaction level model

Similar to other aspects, reliability in large complex electronic systems, e.g. safety-critical CPSs is tackled starting at high level of abstraction. System's fault tolerance is formally checked using UPPAAL and timed automata models generated from AADL specifications [33]. HW design models and tools at such a level also enable verification of interference of extra-functional design aspects [70].

There are research works relying on design soft-error reliability verification by fault-injection campaigns e.g. [69] or formal analysis [68]. This analysis is targeted at extra-functional structures for error protection, e.g. error-correction code (ECC) based mechanisms against single-bit errors in memory elements [68]. [34] proposes a general approach to verify gate-level design transformations for reliability against single-event transients by soft errors that combines formal reasoning on execution traces. [35] and [36] focus on the RAS (reliability, availability and serviceability) group of extra-functional aspects outlined by IBM for complex processor designs where embedded error protection mechanisms and designs intrinsic immunity (due to various masking) to errors is evaluated by fault injection.

[43] and [44] propose extensions to system descriptions in the IP-EXACT format to enable multi-layer representation and simulation of several mutually influencing extra-functional aspects of smart system designs such as lifetime reliability (aging), power and temperature. A complex approach to verification of multiple reliability concerns (soft errors, BTI, etc.) across layers in industrial CPS designs is proposed in [67] as a collaborative research result in the IMMORAL project.

*C. Timing aspects*

Functional temporal properties are essential part of sequential designs' specification that are often modelled for functional verification by computational tree logic (*CTL*), applied for formal approaches, and linear temporal logic (*LTL*) temporal assertions expressed arbitrarily in *PSL* (Property Specification Language), SVA (System Verilog Assertions) or systematically in UVM (Universal Verification Methodology). In the extra-functional domain, these can be extended to specific requirements about *performance* (in particular as a trade-off to the power aspects), *quality of service* parameters such as *latency*, *throughput* etc. For formal classification, particular timing aspects may stay in the "limbo" between functional and extra-functional dimensions when timing properties are indivisible with the functionality for the real-time systems (e.g. demanding a worst-case execution time) [32] or time-constrained communication implementations as in the Network-on-Chip (NoC) structures [37], [38].

Several works have been widely studying system's timing properties. Some researchers are mainly focused on generating timing properties such as Real Time (RT), latency, execution time, throughput, communication time, performance and etc., to reduce the verification process, state space and cost [40], [42], [53]. Other works instead use the timing properties to assess whether the system under verification is correctly functioning or not [41], [50], [52]. In [42] a framework has been developed to analyze performance of a system design. The framework is based on stochastic modeling and simulation and it is applied on a set of NoC topologies. The methodology uses the selective abstraction concept to reduce complexity. [53] introduces a tool called CONTREX to complement current activities in the area of predictable computing platforms and segregation mechanisms with techniques to compute RT properties. CONTREX enables energy efficient and cost aware design through analysis and optimization of properties such as RT. In [41], an analysis tool is developed to work with the AADL [65] developing environment to analyze the latency of the AADL model to assure the correctness of a scheduling model that binds the relation of different components in a model. The authors in [50] modified EASTADL [66] to include energy constraints and transformed energy-aware real-time (ERT) behaviors modeled in EAST-ADL/Stateflow into UPPAAL models amenable to formal verification. And finally, in [52] a platform has been developed to generate a virtual platform in SystemC to express the accuracy of real-time embedded system.

A few works also take into account dependencies between several extra-functional aspects. For instance, the work in [50], [53] and [42] present the effect of optimizing timing properties (performance and latency) on power consumption or the study in [52] performs the effect of decreasing execution time on power consumption. Such analysis is mostly limited to two extra functional aspects or neglected at all [39], [40], [41], [55], while design timing constraints can strongly influence not only power consumption but reliability, security, availability, etc. as well as functional properties.

*D. Power consumption*

This extra-functional aspect has a tight relation to the implementation technology assumed for the synthesis of the design model under verification. With planar bulk MOSFET technology known for exponential growth of the static leakage power for smaller device geometries and employment of FinFET and Tri-Gate-Transistors in the advanced technology nodes, the CMOS device parameters are essential for this analysis [45].

In commercial flows, this verification dimension can be addressed relatively independently from the functional verification dimension. The *power intent* and detailed power modelling can be done starting at TLM or RTL with minimal interference with the HDL functional description, e.g. using the Accellera introduced Unified Power Format (UPF) employed for power-aware design verification automation by commercial tools especially with the latest UPF3.0 [48] or Cadence/Si2 Common Power Format CPF [49]. For the advanced device implementation technologies, power specification implies *multi-voltage design* with up to tens of *power domains* and may consider dynamic and adaptive voltage scaling.

In the recent research works, design verification against the power aspect is performed at different abstraction levels with a trade-off between speed and accuracy. Some works such as [46], [47], [61], [62] perform power analysis at system level targeting high simulation speed and low power optimization flexibility similar to the accuracy achievable at lower levels. In [46], the authors applied their approach to SRAM and AES encryption IPs and obtained a significant simulation speed-up in comparison to gate-level simulation with a high fidelity of the system-level power simulation. A promising software tool for power simulation in SystemC designs is the Powersim framework [47], [62]. In [47], a methodology to estimate the dissipation of energy in hardware at any level of abstraction is proposed. In [62], the authors propose a SystemC class library aimed at calculation of energy consumption of hardware described at system level. The work in [60] introduces a series of tools which can be tightly linked and enable the power analysis from layout, gate-level, RT-level, IP-level to system level. The power aspect verification could benefit from a holistic multi-level modelling, such as e.g. [17] available for functional verification. [42], [43], [44], [50], [52], [53], are aimed at methodologies suitable for specific applications (such as cyber-physical system [50]) that assume verification of extra-functional aspects such as power, timing, thermal at the system level.

## IV. THE CHALLENGE OF MULTIDIMENSIONAL VERIFICATION

The performed analysis of the state of the art has outlined a gap in methodologies and tools for holistic multidimensional verification of hardware design models.

Different from functional verification, approaches for extra-functional hardware design aspects' verification remain underdeveloped even when tackled in isolation. Here, one of the key issues is a lack of established metrics for verification confidence. For a particular functional verification plan, the functional dimension usually includes conventional structural (code) coverage metrics, functional coverage [3] in form of asserted and assumed properties and design parameters along with stimuli quality assessment by model mutations [18]. The metrics for confidence in extra-functional dimension verification results may be challenging as in practice the requirements are *subjective* and can be specified as a mixture of *quantitative* and *qualitative constraints*. Accurate hardware verification in a particular dimension requires both sufficient extra-functional design modeling and the extra-functional aspect target modeling [70]. There is a limited number of dedicated commercial tools and common standards for extra-functional verification flows. In particular, for the security dimension the JasperGold SPV [71] is one of the few such tools that stand out from the academic research frameworks. Finally, the issue of eliciting the extra-functional requirements [4], [5] is a challenging task as ambiguity and (sometimes conflicting) interdependency of the extra-functional aspects in the specifications increases complexity and may leave gaps in the multidimensional verification plans.

Unfortunately, there is no established hardware design methodology supporting multidimensional verification plans for mutually influencing functional and extra-functional aspects. There is a very limited number of research works going beyond analysis of one extra-functional verification aspect under constraints of another as the complexity of the problem grows extremely fast with the number of dimensions (interdependent constraints) and the electronic system size. The first works in this direction are, for example, [44] and [70].

The objective for the research community is to manage multidimensional verification campaigns as illustrated in Fig.2. Fig. 2a is an illustration of six independent verification campaigns in a three-dimensional verification space. Here, a verification campaign can achieve a level of confidence in one, two or all dimensions - (F)unctionality, (P)ower and (S)ecurity. Radar-charts are an instrument for summarizing multidimensional verification results for unlimited number of dimensions, see Fig. 2b (where the dimensions can be ordered to emphasize correlation or interdependencies between adjacent dimensions).

### A. Motivational Example

Single-dimension verification campaigns ignoring interdependencies between the dimensions may lead to gaps in the overall electronic system quality. As an example, let us consider an actual verification campaign of an open-source NoC framework Bonfire [57], [58].

The design under verification is in RTL VHDL and implements a 2x2 NoC infrastructure (processing elements excluded). The verification plan considered 2-dimensional verification campaign targeting *functionality* and *power consumption* requirements. For the former, assertion-based functional verification by simulation was employed targeting statement, branch, condition and toggle coverage metrics and satisfaction of a set of temporal simple-subset PSL assertions. For the latter, a set of power targets were extracted for the targeted silicon implementation assuming a predetermined switching activity.

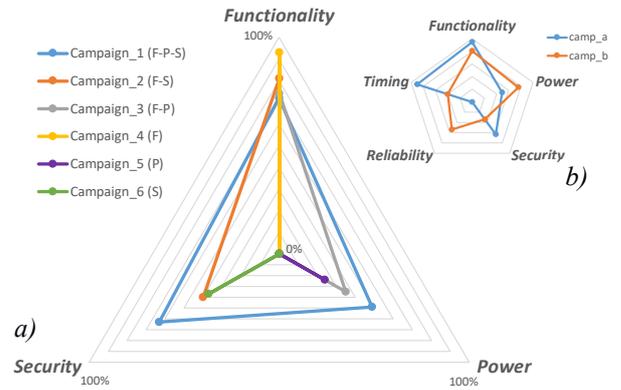

Fig. 2. Multidimensional verification campaigns
(Radar-chart *n*-dimensional visualization)

Among documented design errors, the bug *f1*, as shown in Fig. 3, is an example of a functional misbehavior due to improper usage of write and read pointers in the FIFO. The bug *p1* as shown in Fig. 4, causes violations of specified power consumption targets because of unnecessary excessive use of a fault-tolerance structure related counter. Interestingly, functionality of both the router core and the complete system is not interfered in case of *p1*.

Table II summarizes power consumption for the three cases. Here, the Total Power is composed of the dynamic power, i.e. the Switching Power in the interconnects and the Internal Power in the logic cells, and the insignificant (for the target technology) static leakage power Leak Power. The case *p1* results in double power consumption compared to the correct implementation and violates the power targets in the specification, whereas the power consumption for the *f1 case* remains within the specification. Design verification in a single dimension may lead to a faulty design.

```
process(write_en, write_pointer) begin --write pointer bug
  if write_en = '1' then
    write_pointer_in <= write_pointer(0)&write_pointer(3 downto 1);
  else
    write_pointer_in <= write_pointer;
  end if;
end process;

process(read_en, empty, read_pointer) begin --read pointer bug
  if (read_en = '1' and empty = '0') then
    read_pointer_in <= read_pointer(0)&read_pointer(3 downto 1);
  else
    read_pointer_in <= read_pointer;
  end if;
end process;
```

```
process(write_en, write_pointer )begin --write pointer
  if write_en = '1' then
    write_pointer_in <= write_pointer(2 downto 0)&write_pointer(3);
  else
    write_pointer_in <= write_pointer;
  end if;
end process;

process(read_en, empty, read_pointer) begin --read pointer
  if (read_en = '1' and empty = '0') then
    read_pointer_in <= read_pointer(2 downto 0)&read_pointer(3);
  else
    read_pointer_in <= read_pointer;
  end if;
end process;
```

Fig. 3. Bug *f1* and its correction.

```vhdl
process(Healthy_packet, reset_counters, healthy_counter_out)
begin
  if reset_counters = '1' then
     healthy_counter_in <=  (others => '0');
  elsif Healthy_packet = '1' then             -- Bug p1!
     healthy_counter_in <= healthy_counter_out + 1;
  else
     healthy_counter_in <= healthy_counter_out;
  end if;
end process;
```

```vhdl
process(Healthy_packet, reset_counters, healthy_counter_out,
        faulty_counter_out) begin
  if reset_counters = '1' then
     healthy_counter_in <=  (others => '0');
  elsif Healthy_packet = '1' and faulty_counter_out /=
std_logic_vector(to_unsigned(0, faulty_counter_out'length)) then
     healthy_counter_in <= healthy_counter_out + 1;
  else
     healthy_counter_in <= healthy_counter_out;
  end if;
end process;
```

Fig. 4. Bug *p1* and its correction.

TABLE II. POWER CONSUMPTION OF THE CORRECTED BONFIRE SYSTEM IMPLEMENTATION AND THE ONE IN PRESENCE OF BUGS *F1* AND *P1*.

| Bonfire system Implementation | Switching Power (mW) | Internal Power (mW) | Leak Power (pW) | Total Power (mW) |
|---|---|---|---|---|
| with *f1* bug | 0.783 | 9.427 | 7.50e+05 | 10.211 |
| with *p1* bug | 0.757 | 21.379 | 6.93e+05 | 22.137 |
| corrected | 0.666 | 9.518 | 7.43e+05 | 10.184 |

## V. CONCLUSION

In the recent years, numerous extra-functional aspects of electronic systems were brought to the front and imply verification of hardware design models in multidimensional space along with the functional concerns of the target system. In this paper, we have presented a taxonomy for multidimensional hardware verification aspects, a state-of-the-art survey of related research works and trends towards the multidimensional verification concept. The performed analysis of the state of the art has outlined a gap in methodologies and tools for holistic multidimensional verification of hardware design models. The concept was also motivated by a case study for the functional and power verification dimensions.

## ACKNOWLEDGMENTS

We would like to acknowledge Apneet Kaur and Behrad Niazmand for their help with the case study analysis. This research was supported in part by projects H2020 MSCA ITN RESCUE funded from the EU H2020 programme under the MSC grant agreement No.722325, H2020 TWINN TUTORIAL, by the Estonian Ministry of Education and Research institutional research grant no. IUT19-1 and by European Union through the European Structural and Regional Development Funds.